\documentclass{article}
\usepackage{amsmath}
\usepackage{amssymb}
\usepackage{graphics}
\usepackage{graphicx}
\usepackage[labelsep=period, font=normalsize]{caption}

\begin{document}
\begin{titlepage}

\begin{centering}
\textbf{A New Quantum Interferometric Protocol Using Spin-Dependent Displacements} \\[1.5cm]

\textbf{Necati Çelik\(^{1*}\), Songül Akbulut Özen\(^2\), Burhan Engin\(^2\)} \\[0.5cm]

\(^1\)Gümüşhane University, Faculty of Engineering and Natural Sciences, Department of Physics Engineering, Gümüşhane, Türkiye.\\[0.2cm]

\(^2\)Bursa Technical University, Faculty of Engineering and Natural Sciences, Department of Physics, Türkiye\\[1cm]

\(^*\)Corresponding Author \\[0.2cm]
Email: necati.celik@gumushane.edu.tr\\[0.5cm]

\end{centering}
\end{titlepage}

\section*{Abstract}
We propose a quantum interferometric protocol that leverages spin-dependent spatial displacements to enable high-precision parameter estimation beyond classical limits. By inducing a unitary coupling between a particle’s spin degree of freedom and its momentum, the protocol generates entanglement between spin states and spatial positions, resulting in coherent spatial superpositions. Interferometric reconstruction of the resulting phase differences enables Heisenberg-limited sensitivity for parameters encoded in the spin Hamiltonian. As a concrete application, we demonstrate the protocol’s effectiveness in magnetic field sensing, where the field is transduced into spatial interference fringes. Quantum Fisher information analysis confirms sub-shot-noise scaling, and the protocol's feasibility is discussed for physical platforms including ultracold atoms and nitrogen-vacancy (NV) centers. Our framework provides a versatile approach to quantum metrology with potential extensions to multiparameter sensing and gravitational wave detection.

\vfill

\noindent \textbf{Keywords:} Quantum metrology, Magnetic field sensing, Spin-position entanglement, Spatial superposition.

\section{Introduction}
Precision measurement protocols that surpass classical limits are central to advancing both fundamental science and emerging quantum technologies \cite{Giovannetti2011, Hotter2024}. Quantum metrology offers powerful tools for this purpose by exploiting coherence, entanglement, and superposition \cite{Bohr2024, Yuan2017}. In this work, we introduce a quantum interferometric protocol in which a particle’s internal spin state is coherently coupled to its spatial position via spin-dependent momentum kicks. As a concrete application, we demonstrate how the proposed protocol enables high-precision magnetic field sensing by mapping spin information onto spatial interference patterns. Precise measurement of magnetic fields is a cornerstone of modern science and technology, with applications spanning fundamental physics, medical imaging, navigation systems, and materials characterization \cite{Clarke2006, Kominis2003, Budker2007}. For instance, in biomedical engineering, magnetoencephalography (MEG) relies on ultra-sensitive detection of weak magnetic fields generated by neuronal activity to map brain function non-invasively \cite{Hamalainen1993}. Similarly, in geophysical exploration, high-precision magnetometry enables the detection of mineral deposits and underground structures \cite{Telford1990}. The ability to measure magnetic fields with exceptional sensitivity and resolution thus holds transformative potential across multiple disciplines. Classical magnetometry techniques, such as superconducting quantum interference devices (SQUIDs) or fluxgate sensors, are constrained by the shot-noise limit, where the uncertainty in the measured magnetic field \( B \) scales as \(\Delta B \propto 1/\sqrt{N} \label{eq:shot_noise} \) for \( N \) independent measurements \cite{Giovannetti2004, Degen2017}. Quantum metrology offers a paradigm shift by exploiting nonclassical resources—such as entanglement, superposition, and squeezing—to surpass these classical limits \cite{Pezze2018, Toth2014}. Notably, protocols leveraging entangled states can achieve Heisenberg-limited scaling, where \( \Delta B \propto 1/N \label{eq:heisenberg_limit} \) providing a quadratic improvement in precision \cite{Wineland1992, Giovannetti2011}. However, realizing such enhancements in practical settings remains a significant challenge, particularly in maintaining coherence and mitigating environmental noise.

In this work, we propose a novel quantum metrology protocol that harnesses the entanglement between a particle's internal spin degree of freedom and its external spatial position to achieve sub-shot-noise sensitivity in magnetic field measurements. The main objective of our study is to demonstrate how a unitary coupling between the spin Hamiltonian and the particle's momentum can induce spin-dependent spatial displacements, creating a superposition of position states. By interferometrically reconstructing the resulting spatial interference patterns, we enable magnetic field estimation with precision beyond classical limits. Our approach is distinct from existing methods in several key aspects: 1) Unlike traditional spin-based magnetometers, which rely solely on spin precession, our protocol encodes the magnetic field information into spatial superpositions, enabling direct readout via interferometry. 2) We derive fundamental bounds from quantum Fisher information (QFI) to rigorously establish the \( N \) enhancement for \( N \)-entangled particles, achieving Heisenberg-limited scaling. 3) The protocol is applicable to a wide range of physical systems, from ultracold atoms to solid-state defects, offering flexibility in experimental implementation.

The significance of our work lies in its potential to bridge the gap between theoretical quantum metrology and practical sensing applications. For example, in quantum technologies such as nitrogen-vacancy (NV) centres in diamond, our protocol could enable nanoscale magnetic imaging with unprecedented resolution \cite{Rondin2014}. Similarly, for ultracold atomic systems, it opens avenues for probing weak magnetic fields in exotic quantum phases \cite{Gross2017}.

\section{Materials and Methods}
\subsection{Spin-Position Coupling and Displacement Dynamics}
The foundation of our protocol lies in the controlled entanglement between a particle’s internal spin degree of freedom and its external spatial position. To formalize this coupling, we begin with a spin-\( s \) particle subjected to a magnetic field \( B \), governed by the Zeeman Hamiltonian:

\smallskip
\begin{equation}
\hat{H} = -\gamma B \hat{S}_z
\label{eq:1}
\end{equation}
\smallskip

\noindent
where \( \gamma \) is the gyromagnetic ratio, \(\hat{S}_z \) is the spin operator along the quantization axis, and the eigenvalues \( E_m = -\gamma B \hbar m \) correspond to the spin eigenstates \( |m \rangle \) with \( m = -s, \dots, s \). This Hamiltonian encodes the magnetic field \( B \) into the spin energy levels, with the proportionality to \( m \) reflecting the linear dependence on the spin projection.

To induce the spin-dependent spatial displacements, we introduce a unitary operator that couples the spin Hamiltonian to the particle’s momentum:

\smallskip
\begin{equation}
\hat{U} = e^{-i \frac{k t}{\hbar} \hat{H} \otimes \hat{p}}
\label{eq:2}
\end{equation}
\smallskip

\noindent
where \( k \) is the coupling constant with units of inverse momentum. This unitary operator acts on the joint spin-position Hilbert space, entangling the spin eigenstates \( |m \rangle \) with momentum-dependent spatial translations. Physically, this coupling could be realized through techniques such as magnetic gradient pulses, optical Stark shifts, or spin-orbit interaction, depending on the experimental platform.

To elucidate the action of \( \hat{U} \), consider a particle initially prepared in a spin eigenstate \( |m\rangle \) and localized at position \( x_0 \), represented as the product state \( |m\rangle \otimes |x_0\rangle \). Applying \( \hat{U} \) generates a spin-dependent displacement in position

\smallskip
\begin{equation}
\hat{U} |m\rangle \otimes |x_0\rangle = |m\rangle \otimes e^{-i \frac{k t E_m \hat{p}}{\hbar}}  |x_0\rangle
\label{eq:3}
\end{equation}
\smallskip

The operator \( e^{-i \frac{\Delta x}{\hbar} \hat{p}} \) is the standard translation operator in quantum mechanics, which shifts the spatial wavepacket by \( \Delta x = k E_m \). Substituting \( E_m = -\gamma B \hbar m \), the displacement becomes

\smallskip
\begin{equation}
\Delta x_m = -k t \gamma B \hbar m
\label{eq:4}
\end{equation}
\smallskip

This result demonstrates that each spin eigenstate \( |m\rangle \) is correlated with a distinct spatial displacement proportional to \( m \), thereby encoding the magnetic field \( B \) into the particle’s position. For a spin-\(\frac{1}{2}\) particle (\( m=\pm\frac{1}{2} \)), the displacements are \( \Delta x_{\pm} = \mp \frac{k t \gamma B \hbar}{2} \label{eq:spin_half_displacements} \) creating a superposition of two spatially separated wavepackets.

The displacement dynamics can be generalized to arbitrary spin-\( s \) systems, where the \( 2s+1 \) spin projections result in \( 2s+1 \) distinct spatial shifts. For example, a spin-1 particle (\( m=-1, 0, 1 \)) would exhibit three spatially separated components with displacements \( \Delta x_{-1} = k t \gamma B \hbar \), \(\Delta x_0 = 0 \), \(\Delta x_{+1} = -k t \gamma B \hbar \) . The magnitude of the displacement scales linearly with \( B, k, t, \gamma \), and \( \hbar \), underscoring the role of these parameters in enhancing sensitivity.

The coupling constant \( k \) determines the scale of the displacement per unit magnetic field. Increasing \( k \) enhances sensitivity but must be balanced against practical constraints, such as wavepacket dispersion and decoherence. The gyromagnetic ratio \( \gamma \) is material-dependent; systems with larger \( \gamma \) (e.g., electrons with \( \gamma_e \approx 28 \) GHz/T) enable greater displacements for a given \( B \), compared to nuclei or neutral atoms. The proportionality to \( \hbar \) emphasizes the quantum-mechanical origin of the effect, as classical treatments would lack such displacement superpositions.

\subsection{State Preparation and Entangling Dynamics}

The protocol efficacy hinges on the precise preparation of a spin superposition state and its subsequent entanglement with spatial degrees of freedom. Below, we detail these steps, elucidating the underlying physics and experimental considerations.

The protocol begins by preparing a spin-\( s \) particle in a coherent superposition of eigenstates, localized at an initial position \( x_0 \). The initial state is given by:

\smallskip
\begin{equation}
| \psi_{\text{in}} \rangle = \left( \sum_{m=-s}^{s} C_m |m\rangle \right) \otimes |x_0\rangle
\label{eq:5}
\end{equation}
\smallskip

\noindent
where \( C_m \) are complex coefficients satisfying \( \sum_m |C_m|^2 = 1 \), and \( |x_0\rangle \) represents a spatially localized wavepacket (e.g., Gaussian wavepacket with width \( \sigma \)). For maximal sensitivity, the coefficients \( C_m \) are typically chosen to form a balanced superposition. For instance, in the spin-\(\frac{1}{2}\) case (\( m=\pm\frac{1}{2} \)), the optimal state is

\smallskip
\begin{equation}
| \psi_{\text{in}} \rangle = \frac{1}{\sqrt{2}} \left( |+\rangle + |-\rangle \right) \otimes |x_0\rangle
\label{eq:6}
\end{equation}
\smallskip

\noindent
which maximizes the coherence between spin states and ensures equal contributions to the subsequent interference signal.

Spin superpositions can be prepared using Raman transitions or microwave pulses to coherently drive transitions between spin states using ultracold atoms \cite{Cabedo2022}. In the case of Nitrogen-Vacancy (NV) Centres \cite{Chu2015}, optical pumping and microwave pulses initialize the NV spin into a superposition of \( |m_s = 0\rangle \) and \( |m_s = \pm1\rangle \) states. Spatial confinement can be achieved via optical traps (atoms) or nanoscale positioning (NV centres in diamond).

The prepared spin superposition is then entangled with the particle’s spatial position through the unitary operator \(\hat{ U} \), derived from the spin-momentum coupling Hamiltonian:

\smallskip
\begin{equation}
\hat{U} = e^{-i \frac{k t}{\hbar} \hat{H} \otimes \hat{p}} = e^{\frac{i}{\hbar} k t \gamma B \hat{S}_z \otimes \hat{p}}
\label{eq:7}
\end{equation}
\smallskip

Here, \( k \) quantifies the coupling strength, determined by experimental parameters such as interaction time \( t \) or gradient field strength. For example, in a magnetic field gradient \( \Delta B \), \( k \propto t\Delta B \), translating to larger displacements for stronger or longer interaction times. Applying \(\hat{U}\) to the initial state \( | \psi_{\text{in}} \rangle \) generates spin-dependent spatial displacements:

\smallskip
\begin{equation}
|\Psi\rangle = \hat{U} | \psi_{\text{in}} \rangle = \sum_{m} C_m |m\rangle \otimes |x_0 + \Delta x_m\rangle
\label{eq:8}
\end{equation}
\smallskip

\noindent
where \( \Delta x_m = - k t \gamma B \hbar m \). For spin-\(\frac{1}{2}\) particles

\smallskip
\begin{equation}
|\Psi\rangle = \frac{1}{\sqrt{2}} \left( |+\rangle \otimes |x_0 - \frac{k t \gamma B \hbar}{2}\rangle + |-\rangle \otimes |x_0 + \frac{k t \gamma B \hbar}{2}\rangle \right)
\label{eq:9}
\end{equation}
\smallskip

This entangles the spin states \( |+\rangle \) and \( |-\rangle \) with spatially separated wavepackets. The displacement \( \Delta x_{\pm} = \mp \frac{k t \gamma B \hbar}{2} \) encodes the magnetic field \( B \) into the spatial separation \( \Delta x_+ - \Delta x_- = k t \gamma B \hbar \). The entangling dynamics mirror a quantum-enhanced Stern-Gerlach experiment. In the classical Stern-Gerlach setup, a magnetic gradient separates spin states spatially, but the resolution is limited by shot noise. Here, the unitary \(\hat{ U} \) creates a coherent spatial splitting proportional to \( B \), enabling interferometric readout with Heisenberg-limited precision. The spatial separation \( \Delta x_m \) serves as a “ruler” encoding \( B \), with finer resolution achievable through entanglement.

\begin{figure}[h]
    \centering
    \includegraphics[width=1\linewidth]{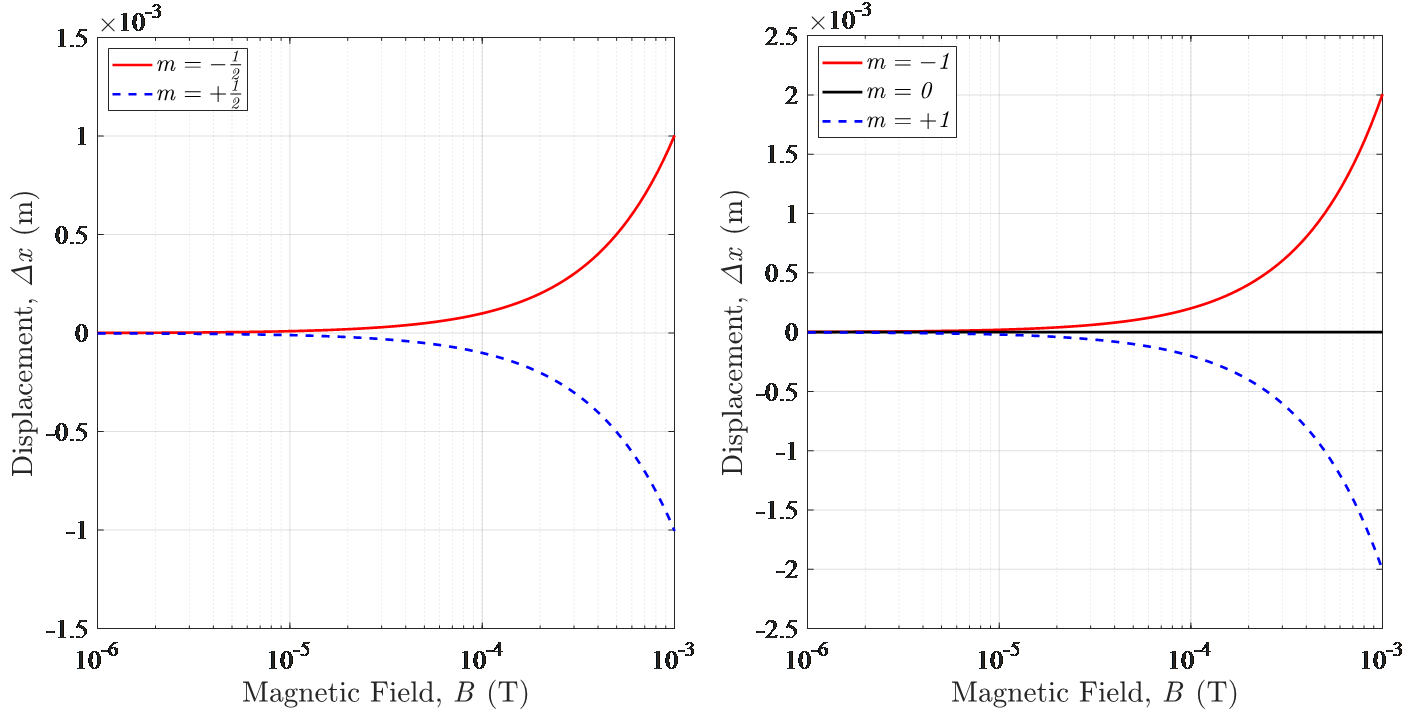}
    \caption{Spin-Dependent Spatial Displacement}
    \label{fig:fig1}
\end{figure}

Figure \ref{fig:fig1} visualizes the protocol’s foundational mechanism: the entanglement between a particle’s spin state and its spatial position, induced by the unitary coupling operator \(\hat{ U} \). For a spin-1/2 particle, the two wavepackets are displaced by \( \Delta x_{\pm} \), forming a coherent superposition separated by a distance proportional to the magnetic field \( B \). For a spin-1 particle, the figure shows three wavepackets corresponding to spin projections \( m=0 \) and \( m=\pm1 \). This increased spatial separation facilitates more distinguishable interference patterns, thereby improving the visibility and precision of interferometric magnetic field measurements. These displacements exemplify how higher-spin systems produce richer spatial structures and enable more complex interference patterns.

\subsection{Interferometric Measurement}

Following the entangling dynamics, the system undergoes free evolution to enable interferometric measurement of the magnetic field \( B \). Each displaced spatial state \( |x_0 + \Delta x_m\rangle \) evolves under the free particle Hamiltonian \(\hat{H}_{\text{free}} = \frac{\hat{p}^2}{2 m_p} \) where \( m_p \) is the particle mass. The time-evolved wavefunction for a spin eigenstate \( |m\rangle \) is obtained via the unitary operator \(\hat{U}_{\text{free}} = e^{- \frac{i}{\hbar} \hat{H}_{\text{free}} t } \). For an initial Gaussian wavepacket centered at \( x_0 + \Delta x_m \) with width \( \sigma \), the time-evolved wavefunction becomes:

\smallskip
\begin{equation}
\tiny \psi_m(x,t) \!=\! \left( 2 \pi \sigma_t^2 \right)^{-1/4} \!exp\! \left(- \frac{(x - x_0 - \Delta x_m)^2}{4 \sigma_t^2} \right) 
\!exp\! \left(i \frac{m_p (x - x_0 - \Delta x_m)^2}{2 \hbar t} - \frac{i}{2} \!arctan\! \left(\frac{\hbar t}{2 m_p \sigma^2} \right) \right)
\label{eq:10}
\end{equation}
\smallskip

\noindent
where \( \sigma_t = \sigma \sqrt{1 + \left(\frac{\hbar t}{2 m_p \sigma^2} \right)^2} \) describes wavepacket spreading \cite{Huber1987}.

\begin{figure}[h]
    \centering
    \includegraphics[width=0.8\linewidth]{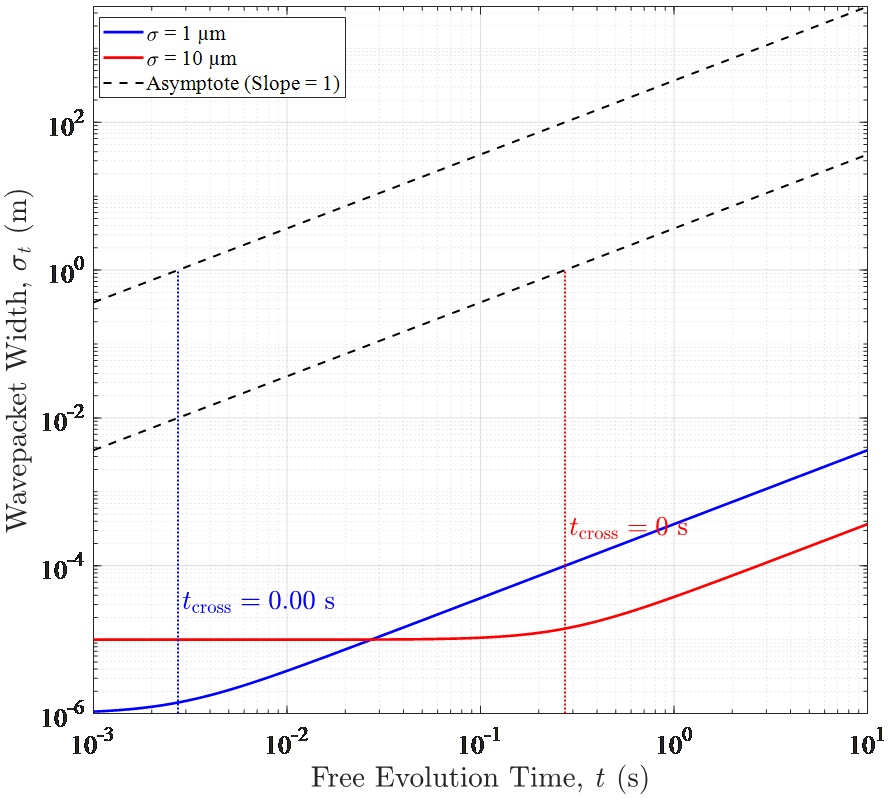}
    \caption{Wavepacket Dispersion over Time}
    \label{fig:fig2}
\end{figure}

As depicted in Figure \ref{fig:fig2}, wavepacket dispersion grows with time, reducing spatial overlap between displaced states. This effect limits the maximum effective evolution time and imposes a trade-off between sensitivity and interference visibility.

The phase term includes a quadratic spatial dependence and a time-dependent contribution from dispersion. The term \( \exp \left(i \frac{m_p (x - x_0 - \Delta x_m)^2}{2 \hbar t} \right) \) corresponds to the kinetic phase acquired due to free propagation, the term \small {\( \exp \left(\frac{i}{2} \!arctan\! \left(\frac{\hbar t}{2 m_p \sigma^2} \right) \right) \)} is the quantum analog of the Gouy phase \cite{Marinho2024} for matter waves, arising from wavepacket spreading. The total wavefunction after free evolution is a superposition of spin-dependent spatial states;

\smallskip
\begin{equation}
\Psi(x,t) = \sum_{m} C_m \psi_m(x,t)
\label{eq:11}
\end{equation}
\smallskip

where, \( C_m \) are spin-state amplitudes, and \( \psi_m(x,t) \) are time-evolved spatial wavefunctions for each spin projection \( m \). Eq.\ref{eq:11} is obtained after applying the unitary spin-position coupling (Eq.\ref{eq:8}) and free evolution (Eq.\ref{eq:10}). The resulting total state is weighted sum of individual spin-dependent wavepackets. For a spin-1/2 particle (\( m=\pm \frac{1}{2}\)), this reduces to

\smallskip
\begin{equation}
\Psi(x,t) = C_+ \psi_+(x,t) + C_- \psi_-(x,t)
\label{eq:12}
\end{equation}
\smallskip

\noindent
where \(\psi_\pm(x,t) \) corresponds to displacements \(\Delta x_\pm = \mp \frac{k t \gamma B \hbar}{2} \). The probability density \( |\Psi(x,t)|^2 \) contains interference terms:

\smallskip
\begin{equation}
|\Psi(x,t)|^2 = |C_+|^2 |\psi_+(x,t)|^2 + |C_-|^2 |\psi_-(x,t)|^2 + 2 \text{Re} \left[ C_+ C_-^* \psi_+(x,t) \psi_-^*(x,t) \right]
\label{eq:13}
\end{equation}
\smallskip

In Eq.\ref{eq:13}, while \( |C_+|^2 |\psi_+|^2 \) and \( |C_-|^2 |\psi_-|^2 \) represent classical probabilities, \newline \( 2 \text{Re} \left[ C_+ C_-^* \psi_+\psi_-^*\right] \) encodes quantum interference. This term is critical for detecting phase differences between spin states, which encode magnetic field \( B \) into spatial fringes. Interference requires particle overlap between \( \psi_+(x,t) \) and \( \psi_-(x,t) \). For maximal sensitivity, \( |C_+| = |C_-| = \frac{1}{\sqrt{2}} \) ensures equal weighting of spin states and optimal fringe contrast. Additionally, for maximum visibility, the condition \( \Delta x_+ - \Delta x_- \sim \sigma_t \) should be satisfied \cite{Romero-Isart2011}. The interference term scales as \( C_+ C_-^* \), enabling Heisenberg-limited sensitivity when entangled states (e.g., GHZ) are used. Without coherence (\( C_+ C_-^* = 0 \)), the uncertainty reduces to shot-noise scaling \cite{Pezze2014} \( \Delta B \propto 1/\sqrt{N} \label{eq:shot_noise_scaling} \). The interference term depends on the relative phase

\smallskip
\begin{equation}
\Delta \Phi = \Phi_+ - \Phi_-
\label{eq:14}
\end{equation}
\smallskip

\noindent
where \( \Phi_{\pm} \) are the phases of \( \psi_{\pm}(x,t) \) in Eq.\ref{eq:10}. Substituting the displacements \( \Delta x_{\pm} \), the phase difference becomes

\smallskip
\begin{equation}
\Delta \Phi = m_p k \gamma B (x - x_0)
\label{eq:15}
\end{equation}
\smallskip

\(\Delta \Phi \) arises from the distinct phases \(\Phi_\pm\) of the spin-up and spin-down wavepackets, which encode the magnetic field \(B \) into spatial interference pattern. The linear dependence of \( \Delta \Phi \) on \( (x - x_0) \) generates spatially periodic fringes, with spacing proportional to \( B^{-1} \). The fringe spacing \( \lambda_{\text{fringe}} \) is derived from \( \Delta \Phi = 2\pi \frac{(x - x_0)}{\lambda_{\text{fringe}}} \), yielding

\smallskip
\begin{equation}
\lambda_{\text{fringe}} = \frac{2\pi \hbar}{m_p k \gamma B}
\label{eq:16}
\end{equation}
\smallskip

\begin{figure}[h]
    \centering
    \includegraphics[width=0.8\linewidth]{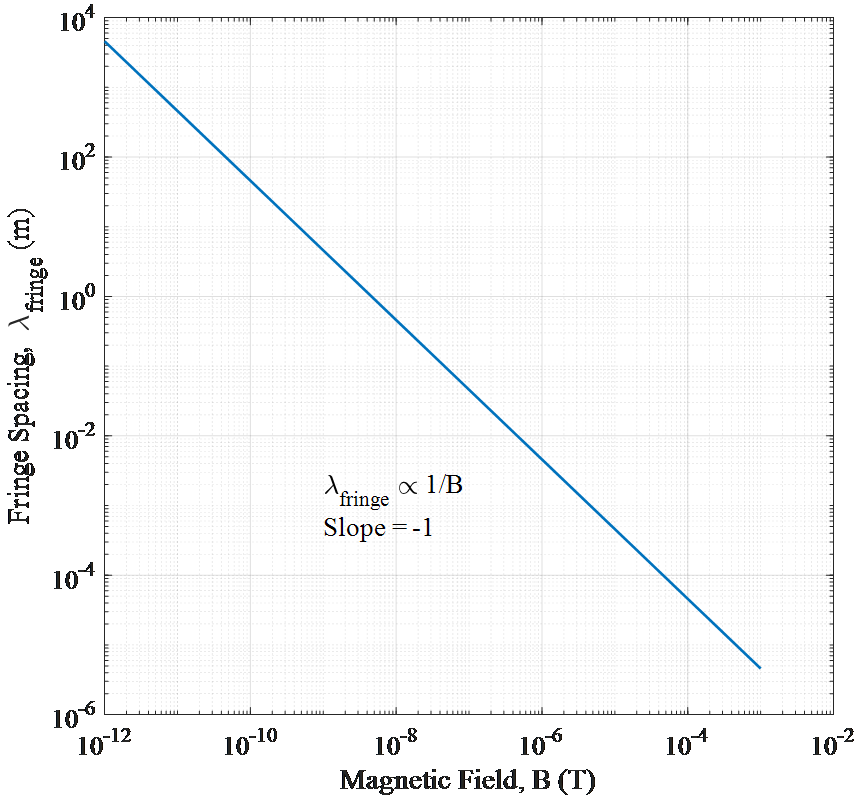}
    \caption{Interference Fringe Spacing vs Magnetic Field}
    \label{fig:fig3}
\end{figure}

The fringe spacing shown in Figure \ref{fig:fig3} decreases with increasing magnetic field \( B \), enabling high-resolution measurements. This inverse relationship is a direct consequence of the magnetic-field-induced phase difference in Eq.\ref{eq:15}.

The measured phase \( \phi \) at a position \( x \) is proportional to the accumulated phase difference:

\smallskip
\begin{equation}
\phi = \frac{\Phi}{2\pi} = \frac{m_p k \gamma B (x - x_0)}{2\pi}
\label{eq:17}
\end{equation}
\smallskip

Using the de Broglie wavelength \( \lambda_{\text{dB}} = \frac{h}{m_p v} \), where \( v = \frac{(x - x_0)}{t} \), this simplifies to

\smallskip
\begin{equation}
B = \frac{\phi \lambda_{\text{dB}} }{k \gamma\hbar t}
\label{eq:18}
\end{equation}
\smallskip

The fringe visibility \(\nu \propto \exp \left( - \frac{(\Delta x_+ - \Delta x_-)^2}{8 \sigma_t^2} \right) \label{eq:fringe_visibility} \) where \( |\Delta x_+ - \Delta x_-| = k t \gamma B \hbar \). High visibility requires \( k \gamma B \hbar t \leq \sigma_t \).

\begin{figure}[h]
    \centering
    \includegraphics[width=0.8\linewidth]{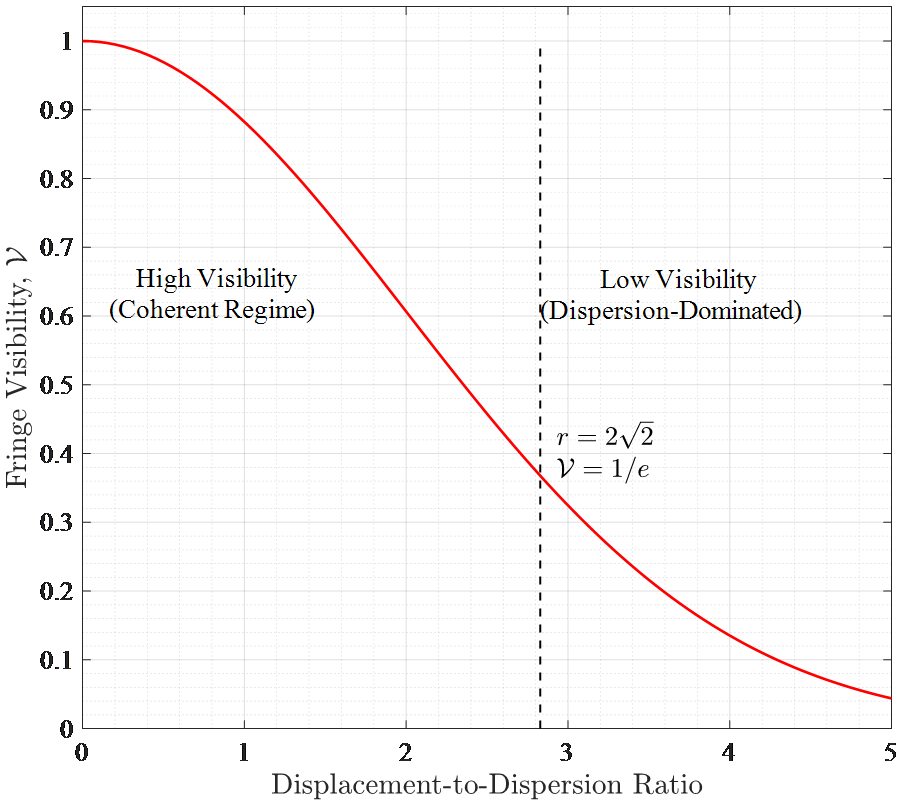}
    \caption{Visibility vs Displacement-to-Dispersion Ratio}
    \label{fig:fig4}
\end{figure}

As illustrated in Figure \ref{fig:fig4}, interference visibility drops sharply when the displacement exceeds the dispersion width \(\sigma_t \). Maintaining \(\Delta x < \sigma_t \) is crucial for ensuring measurable interference fringes and optimal sensitivity. 
The uncertainty in \( B \) is determined by the fringe resolution \( \Delta \phi \). From \( \phi = \frac{m_p k \gamma B (x - x_0)}{2\pi } \), the sensitivity is

\smallskip
\begin{equation}
\frac{d\phi}{dB} = \frac{m_p k \gamma (x - x_0)}{2\pi}
\label{eq:19}
\end{equation}
\smallskip

Using \( \Delta B = \frac{\Delta \phi}{d\phi/dB} \), we find \(\Delta B = \frac{2\pi \Delta \phi}{m_p k \gamma (x - x_0)}
\label{eq:sensitivity_equation} \). Substituting \( m_p (x - x_0) = \frac{h t}{\lambda_{\text{dB}}} \), we find

\smallskip
\begin{equation}
\Delta B = \frac{2\pi\Delta \phi \lambda_{\text{dB}}}{k \gamma h t} = \frac{\Delta \phi \lambda_{dB}}{k \gamma \hbar t}
\label{eq:20}
\end{equation}
\smallskip

\begin{figure}[h]
    \centering
    \includegraphics[width=0.8\linewidth]{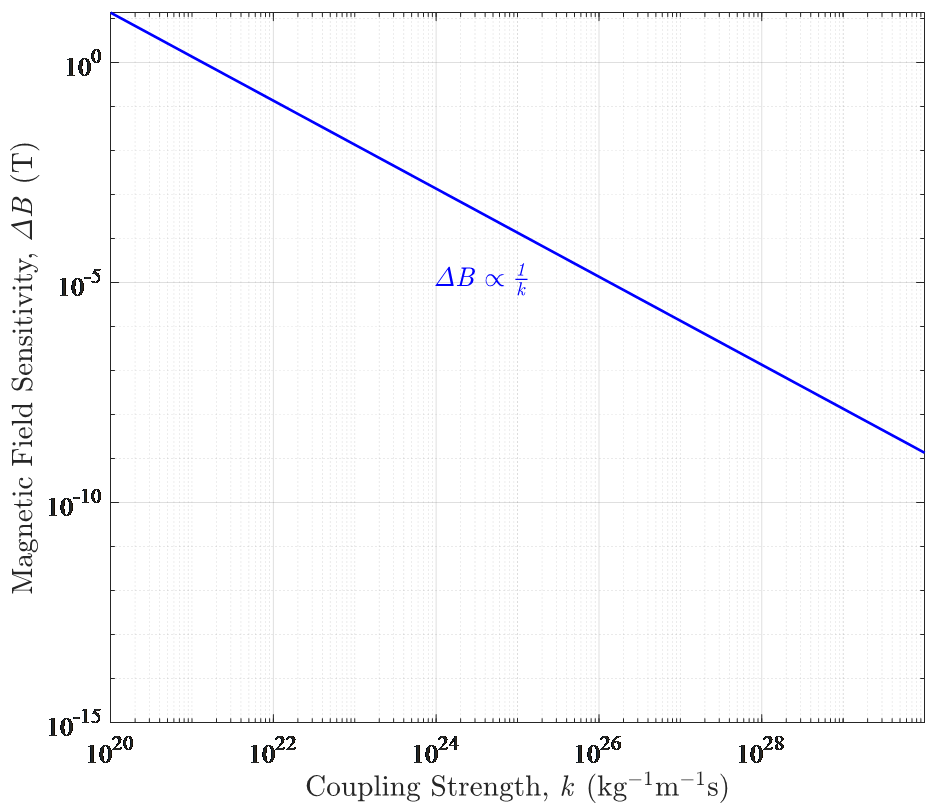}
    \caption{Magnetic Field Sensitivity vs Coupling Strength}
    \label{fig:fig5}
\end{figure}

Figure \ref{fig:fig5} shows the sensitivity \(\Delta B \) as a function of the coupling constant \(k\) revealing an inverse relationship. While larger \(k\) enhances sensitivity by increasing displacement, excessive values risk diminishing interference visibility due to dispersion. 

\section{Results}
\subsection{Quantum Fisher Information}

The maximum precision of \( B \) is dictated by the quantum Fisher information (QFI) \(\mathfrak{T}_Q \). For a spin-1/2 particle, the state after entangling operation \( \hat{U} \) is

\smallskip
\begin{equation}
|\Psi(B) \rangle = C_+ |+\rangle \otimes |x_0 + \Delta x_+ \rangle + C_- |-\rangle \otimes |x_0 + \Delta x_- \rangle
\label{eq:21}
\end{equation}
\smallskip

\noindent
where \( \Delta x_{\pm} = \mp \frac{k t \gamma B \hbar}{2} \) depends on \( B \). For pure states, the QFI reduces to

\smallskip
\begin{equation}
\mathfrak{T}_Q = 4 \left( \langle \partial_B \Psi(B) | \partial_B \Psi(B) \rangle - |\langle \Psi(B) | \partial_B \Psi(B) \rangle|^2 \right)
\label{eq:22}
\end{equation}
\smallskip

The derivative of the state \( |\Psi(B) \rangle \) with respect to \( B \) is

\smallskip
\begin{equation}
|\partial_B \Psi(B) \rangle =  C_+ |+\rangle \otimes \partial_B |x_0 + \Delta x_+ \rangle + C_- |-\rangle \otimes \partial_B |x_0 + \Delta x_- \rangle 
\label{eq:23}
\end{equation}
\smallskip

\noindent
\(\partial_B | x_0 + \Delta x_\pm  \rangle = (\partial_B \Delta x_\pm)(\partial_{\Delta x_\pm}|x_0 + \Delta x_\pm \rangle)   \), and noting that \(\partial_B \Delta x_\pm = \mp\frac{k t \gamma \hbar}{2}\) we get

\smallskip
\begin{equation}
    | \partial_B \Psi(B) \rangle = - \frac{k t \gamma \hbar}{2} (C_+ |+ \rangle \otimes \partial_{\Delta x_+}|x_0 + \Delta x_+ \rangle - C_- |- \rangle \otimes \partial_{\Delta x_-}|x_0 + \Delta x_- \rangle )
    \label{eq:24}
\end{equation}
\smallskip

After evaluating the inner product, the term \(\langle \partial_B \Psi(B)|\partial_B \Psi(B)\rangle\) becomes

\smallskip
\begin{equation}
    (\frac{kt\gamma\hbar}{2})^2\ (|C_+ |^2 \langle \partial_{\Delta x_+} x_0 + \Delta x_+ | \partial_{\Delta x_+}x_0 + \Delta x_+ \rangle + |C_-|^2 \langle \partial_{\Delta x_-}x_0 + \Delta x_- |\partial_{\Delta x_-}x_0 + \Delta x_- \rangle )
    \label{eq:25}
\end{equation}
\smallskip

For a wavepacket \(|x_0 + \Delta x_\pm \rangle\), the derivative \(\partial_{\Delta x_\pm}\) is proportional to the momentum operator \(\hat{p}\), and its expectation value scales with the wavepacket’s momentum spread. For simplicity, if the wavepacket is well-localized, then \(\langle \partial_{\Delta x_\pm} x_0 + \Delta x_\pm |\partial_{\Delta x_\pm} x_0 + \Delta x_\pm \rangle \approx 1\). Then

\smallskip
 \begin{equation}
     \langle \Psi(B)|\partial_B \Psi(B) \rangle = - \frac{k t \gamma \hbar}{2} (|C_+|^2 \langle x_0 + \Delta x_+ | \partial_{\Delta x_+} x_0 + \Delta x_+ \rangle - |C_-|^2 \langle x_0 + \Delta x_- |\partial_{\Delta x_-} x_0 + \Delta x_- \rangle)
     \label{eq:26}
 \end{equation}
 \smallskip

For symmetric wavepackets, \(\langle  x_0 + \Delta x_\pm |\partial_{\Delta x_\pm} x_0 + \Delta x_\pm \rangle = 0\) since the derivative is odd about the centre. After combining terms, the QFI becomes
\smallskip
\begin{equation}
\mathfrak{T}_Q = ( k \gamma \hbar)^2 (|C_+|^2 + |C_-|^2 - 0)=(\frac{k t \gamma \hbar}{2 \sigma})^2
\label{eq:27}
\end{equation}
\smallskip

\noindent
where \(|C_+|^2 + |C_-|^2 = 1\) for normalization. The uncertainty in \( B \) then scales as

\smallskip
\begin{equation}
\Delta B \geq \frac{1}{\sqrt{\mathfrak{T}_Q}} = \frac{2 \sigma}{k t \gamma \hbar}
\label{eq:28}
\end{equation}
\smallskip

\begin{figure}[h]
    \centering
    \includegraphics[width=0.8\linewidth]{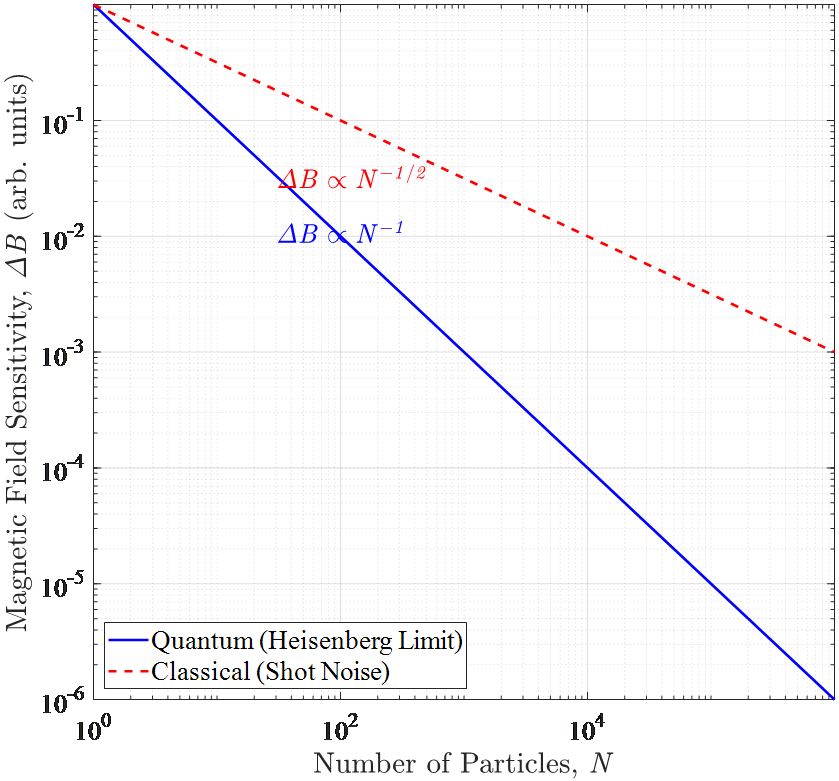}
    \caption{Sensitivity Scaling: Quantum vs Classical}
    \label{fig:fig6}
\end{figure}

As shown in Figure \ref{fig:fig6}, the uncertainty in magnetic field estimation \(\Delta B\) decreases more rapidly with the number of particles \(N\) in the quantum-enhanced protocol (Heisenberg scaling) than in the classical (shot-noise-limited) scenario. This highlights the quantum advantage achieved through entanglement, as established in Eqs.\ref{eq:28} and \ref{eq:35}. The figure thus visualizes the transition from shot-noise-limited sensitivity to Heisenberg-limited precision, underpinned by the quantum Fisher information (QFI) analysis in Eqs.\ref{eq:27} and \ref{eq:34}. The \(N^2\) scaling of the QFI for entangled systems directly enables this enhancement, demonstrating the critical role of entanglement in surpassing classical limits.

\subsection{N-Particle Entangled State}

The protocol’s sensitivity can be further enhanced by employing multiparticle entangled states, which exploit quantum correlations to achieve Heisenberg-limited scaling. Here, we extend the framework to \( N \) spin-1/2 particles prepared in a Greenberger-Horne-Zeilinger (GHZ) state \cite{Berrada2012}, a maximally entangled state of the form

\smallskip
\begin{equation}
|\Psi_{\text{GHZ}} \rangle = \frac{1}{\sqrt{2}} \left( |+\rangle^{\otimes N} + |-\rangle^{\otimes N} \right) \otimes |x_0\rangle^{\otimes N}
\label{eq:29}
\end{equation}
\smallskip

\noindent
where \( |+\rangle \) and \( |-\rangle \) denote the spin eigenstates \( m=+1/2 \) and \( m=-1/2 \), respectively. The GHZ state encodes a coherent superposition of all \( N \) spins aligned parallel or antiparallel to the quantization axis, maximizing the entanglement-enhanced sensitivity to the magnetic field \( B \).

Preparing GHZ states requires precise control over spin-spin interactions and collective operations. For ultracold atoms, this can be achieved via controlled collisions or cavity-mediated entanglement \cite{Eltony2016}. In solid-state systems like nitrogen-vacancy (NV) centers, microwave pulses combined with dynamical decoupling sequences can generate multipartite entanglement \cite{Zhao2013}. After initialization, the unitary operator \(\hat{U} \) acts collectively on all \( N \) particles, entangling their spin states with spatial displacements. The joint unitary operation becomes

\smallskip
\begin{equation}
\hat{U}_{\text{total}} = \prod_{j=1}^{N} e^{\frac{i}{h} k t \gamma B \hat{S}_z^{(j)} \otimes \hat{p}^{(j)}}
\label{eq:30}
\end{equation}
\smallskip

Here multiplication is tensor product \(\hat{S}_z^{(j)} \) and \(\hat{p}^{(j)} \) are the spin and momentum operators for the \( j \)-th particle. Applying \(\hat{U}_{\text{total}} \) to the GHZ state results in

\smallskip
\begin{equation}
|\Psi_{\text{GHZ}} \rangle = \frac{1}{\sqrt{2}} \left( |+\rangle^{\otimes N} \otimes |x_0 + \Delta x_+ \rangle^{\otimes N} + |-\rangle^{\otimes N} \otimes |x_0 + \Delta x_- \rangle^{\otimes N} \right)
\label{eq:31}
\end{equation}
\smallskip

\noindent
where \( \Delta x_{\pm} = \mp \frac{k t \gamma B \hbar}{2} \). Crucially, the displacement for each spin state scales collectively, leading to a total spatial separation proportional to \( N \Delta x_{\pm} \).

The QFI for the GHZ state is derived by computing the sensitivity of \( |\Psi_{\text{final}} \rangle \) to variations in \( B \). For a pure state, the QFI is given by

\smallskip
\begin{equation}
\mathfrak{T}_Q = 4 \left( \langle \partial_B \Psi_{\text{final}} | \partial_B \Psi_{\text{final}} \rangle - |\langle \Psi_{\text{final}} | \partial_B \Psi_{\text{final}} \rangle|^2 \right)
\label{eq:32}
\end{equation}
\smallskip

The derivative of the state with respect to \( B \) is

\smallskip
\begin{equation}
|\partial_B \Psi_{\text{final}} \rangle = \frac{N k t \gamma \hbar}{2 \sqrt{2}} \left( |+\rangle^{\otimes N} \otimes \partial_{\Delta x_+} |x_0 + \Delta x_+ \rangle^{\otimes N} - |-\rangle^{\otimes N} \otimes \partial_{\Delta x_-} |x_0 + \Delta x_- \rangle^{\otimes N} \right)
\label{eq:33}
\end{equation}
\smallskip

Assuming spatially localized wavepackets with negligible overlap, the inner product \( \langle \Psi_{\text{final}} | \partial_B \Psi_{\text{final}} \rangle \) vanishes due to orthogonality. The remaining term yields

\smallskip
\begin{equation}
\mathfrak{T}_Q = N^2 \left( \frac{k t \gamma \hbar}{4 \sigma} \right)^2
\label{eq:34}
\end{equation}
\smallskip

This quadratic scaling with \( N \) arises from the collective enhancement of the displacement in the GHZ state. Consequently, the uncertainty in \( B \) becomes

\smallskip
\begin{equation}
\Delta B \geq \frac{1}{\sqrt{\mathfrak{T}_Q}} = \frac{4 \sigma}{N k t \gamma \hbar}
\label{eq:35}
\end{equation}
\smallskip

\noindent
demonstrating Heisenberg-limited scaling.

It should be noted that GHZ states are highly susceptible to decoherence, as the loss of a single particle collapses the entire superposition. For spatial superpositions, wavepacket dispersion (\(\sigma_t \propto \sqrt{t}\)) and environmental noise (e.g., magnetic field fluctuations) further degrade fringe visibility. To overcome these challenges, one can employ dynamical decoupling using spin-echo sequences to suppress low-frequency noise \cite{Abobeih2022}, minimize free evolution time to reduce dispersion \cite{Choi2023}, and implement error-corrected sensing through decoherence-free subspaces or redundant encoding to preserve entanglement \cite{Harrington2022}. For experimental realization, ultracold atoms in optical lattices offer long coherence times (\(>1\) s) and scalability to large \(N\), though strong spin-momentum coupling requires intense magnetic gradients or synthetic gauge fields \cite{Schafer2020}, while diamond NV centers allow room-temperature operation with nanoscale spatial resolution, and despite challenges in generating entanglement for \( N > 10 \), recent progress in spin-photon interfaces shows promise \cite{Schirhagl2014}. The scalability of the protocol is limited by several factors: GHZ states decohere at a rate proportional to \( \sim N \Gamma \), where \( \Gamma \) is the single-particle decoherence rate; imperfections in state preparation and readout accumulate with increasing \( N \); and for massive particles, wavepacket dispersion restricts the interrogation time \( t \).

\section{Discussion}
The proposed quantum metrology protocol demonstrates a significant advancement in magnetic field sensing by leveraging the entanglement between a particle’s spin and its spatial degree of freedom. The protocol achieves a Heisenberg-limited scaling of precision (\(\Delta B \propto 1/N\)) for \(N\) entangled particles, surpassing the classical shot-noise-limit (\(\Delta B \propto 1/\sqrt{N}\)). This enhancement arises from the coherent superposition of spin eigenstates, which allows simultaneous probing of the magnetic field through all \(2s+1\) spin projections. The \(N^2\) scaling of the QFI for GHZ states highlights the critical role of entanglement. Without entanglement, the best achievable scaling is linear in \(N\). Classical magnetometry relies on independent measurements, limiting sensitivity to \(\Delta B \sim 1/\sqrt{N}\).

The quantum protocol exploits interference between spatially displaced wavepackets, enabling sub-shot-noise precision. While the theoretical framework we proposed is promising, practical implementation faces several challenges. The free evolution of spatial states (Eq.\ref{eq:8}) leads to wavepacket dispersion (\(\sigma_t \propto t\)), which reduces fringe visibility \(\nu\). High precision requires \(|\Delta x_+ - \Delta x_-| \leq \sigma_t\), constraining the coupling strength \(k\) or measurement time \(t\). Environmental noise (e.g., thermal fluctuations or stray magnetic field) can disrupt spin-position entanglement. Techniques such as dynamical decoupling or cryogenic isolation may be necessary \cite{Keil2021}.

For experimental realizations, ultracold atoms \cite{Jaksch2004} offer long coherence times and precise control via optical traps, but require sophisticated laser systems. Nitrogen-Vacancy Centres \cite{Liu2019} provide room-temperature operation and high spin coherence but face challenges in achieving large spatial superpositions.

\begin{figure}[h]
    \centering
    \includegraphics[width=0.8\linewidth]{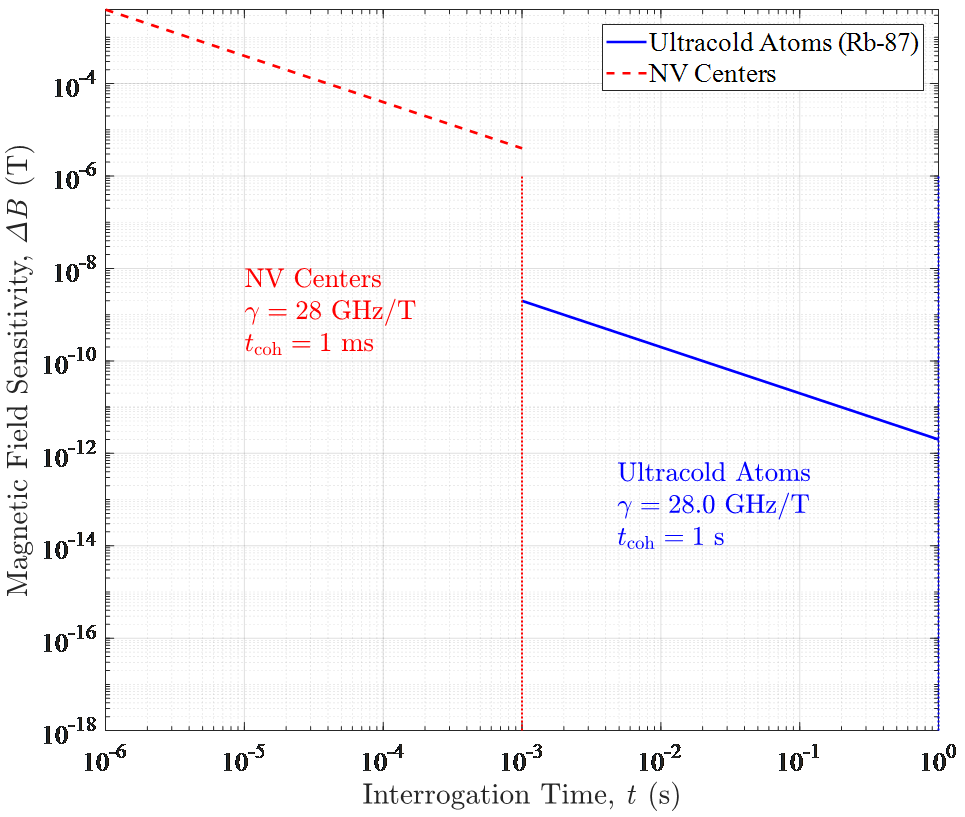}
    \caption{Platform Comparison: Ultracold Atoms vs NV Centres}
    \label{fig:fig7}
\end{figure}

Figure \ref{fig:fig7} compares two key experimental platforms—ultracold atoms and NV centres—for implementing the protocol. It highlights the trade-offs between coherence time, spatial resolution, entanglement scalability, and technical constraints. For ultracold atoms, long coherence times (\(>1\) s) and precise spatial control via optical lattices enable robust entanglement generation (e.g., GHZ states via controlled collisions or cavity-mediated interactions, Eq.\ref{eq:29}-\ref{eq:30}) and reduced dispersion due to larger particle mass \(m_p\). These features favour scalable Heisenberg-limited sensitivity. Conversely, NV centres offer room-temperature operation and nanoscale spatial resolution but face challenges in generating large spatial displacements. Their high gyromagnetic ratio enhances displacement per unit \(B\), yet rapid decoherence and wavepacket dispersion limit practical scalability. The figure underscores the trade-off between coherence time (favouring atoms) and spatial resolution (favouring NV centres), with ultracold atoms excelling in parameter regimes requiring large \(N\) entanglement and NV centres prioritizing nanoscale applications. Both platforms must balance coupling strength \(k\), interrogation time \(t\), and environmental noise to optimize \(\Delta B\).

The uncertainty in \(B\) (Eq.\ref{eq:24}) depends on coupling constant (\(k\)), gyromagnetic ratio (\(\gamma\)), particle mass (\(m_p\)). Larger \(k\) enhances sensitivity but must balance against experimental constraints (e.g., wavepacket width \(\sigma\)). Systems with larger gyromagnetic ratio offer greater displacement per unit \(B\). Lighter particles (e.g., electrons) exhibit larger spatial displacements for a given \(B\), but heavier particles (e.g., atoms) may have slower dispersion. Increasing coupling constant and gyromagnetic ratio improves precision but may exacerbate decoherence or technical noise. Optimal parameters must be determined for specific experimental platforms.

The protocol exemplifies how quantum resources can transcend classical limits in sensing. While challenges remain in experimental realization, the theoretical framework provides a clear pathway toward sub-shot-noise magnetometry and beyond.

\section{Conclusion}
The quantum metrology protocol presented in this work represents a significant advancement in high-precision magnetic field sensing by harnessing the unique properties of quantum mechanics (superposition, entanglement, and spatial interferometry). By coupling a particle’s spin degree of freedom to its spatial position through a unitary interaction, we have demonstrated a theoretically robust framework for achieving Heisenberg-limited precision, surpassing the classical shot-noise limit. The protocol encodes the magnetic field \( B \) into spatial displacement of a particle’s wavepacket, leveraging spin-dependent superpositions. The resulting interference fringes provide a direct measure of \( B \) with precision scaling as \( \Delta B \propto 1/N \) for \( N \)-particle entangled states, a quadratic improvement over classical method (\( \Delta B \propto 1/\sqrt{N} \)). The derivation of the QFI for spin-1/2 and GHZ states rigorously establishes the protocol’s fundamental limits, confirming its superiority in parameter estimation. The QFI’s \( N^2 \) scaling for entangled systems underscores the critical role for nonclassical correlations.

While the protocol is theoretically general, we identified two promising experimental platforms: ultracold atoms which offer long coherence times and precise control but require challenging laser-cooling techniques and Nitrogen-Vacancy Centres which enable room-temperature operations but face limitations in achieving large spatial separations.

We have demonstrated a quantum metrology protocol where Hamiltonian eigenvalues are transduced into spatial superpositions, enabling parameter estimation with precision bounded only by quantum mechanics. Extensions to multiparameter sensing and gravitational wave detection are promising future directions.

\end{document}